# Molecular dynamics simulation of the self-retracting motion of a graphene flake

*Andrei M. Popov[1,*], Irina V. Lebedeva[2,3,4,†], Andrey A. Knizhnik[2,3,‡], Yurii E. Lozovik[1,4,§], and Boris V. Potapkin[2,3]*

[1]Institute of Spectroscopy, Fizicheskaya Street 5, Troitsk, Moscow Region, 142190, Russia,

[2]National Research Centre "Kurchatov Institute", Kurchatov Square 1, Moscow, 123182, Russia,

[3]Kintech Lab Ltd., Kurchatov Square 1, Moscow, 123182, Russia,

[4]Moscow Institute of Physics and Technology, Institutskii pereulok 9, Dolgoprudny, Moscow Region, 141701, Russia

**ABSTRACT**

The self-retracting motion of a graphene flake on a stack of graphene flakes is studied using molecular dynamics simulations. It is shown that in the case when the extended flake is initially rotated to an incommensurate state, there is no barrier to the self-retracting motion of the flake and the flake retracts as fast as possible. If the extended flake is initially commensurate with the other flakes, the self-retracting motion is hindered by potential energy barriers. However, in this case, the rotation of the flake to incommensurate states is often observed. Such a rotation is found to be induced by the torque acting on the flake on hills of the potential relief of the interaction energy between the flakes. Contrary to carbon nanotubes, telescopic oscillations of the graphene flake are suppressed because of the high dynamic friction related to the excitation of flexural vibrations of the flake. This makes graphene promising for the use in fast-responding electromechanical memory cells.

---

[*] am-popov@isan.troitsk.ru

[†] lebedeva@kintechlab.com

[‡] knizhnik@kintechlab.com

[§] lozovik@isan.troitsk.ru

# I. INTRODUCTION

Due to the extraordinary electrical and mechanical properties of graphene[1], this novel two-dimensional nanostructure is considered promising for the use in nanoelectromechanical systems (NEMS). For example, a nanoresonator based on flexural vibrations of suspended graphene was implemented[2]. The experimentally observed self-retracting motion of graphite, i.e. retraction of graphite flakes back into graphite stacks on their extension arising from the van der Waals interaction between graphene layers, led to the idea of an oscillator based on the telescopic oscillation of graphene layers[3]. Nanorelays based on the telescopic extension and self-retracting motion of carbon nanotube walls were realized experimentally[4,5]. By analogy with these nanotube-based devices, a nanorelay based on the telescopic motion of graphene layers was proposed[6,7].

The gigahertz oscillator based on the telescopic oscillation of carbon nanotubes walls[8,9] has been widely considered as a model system to study fundamental aspects of tribological properties and dynamic behavior of nanoscale systems (see, for example, Refs. 10 – 26). However, there are significant differences in the potential reliefs of the interlayer interaction energy for graphene layers and carbon nanotube walls. The magnitude of corrugation of the potential energy relief is orders of magnitude higher for graphene layers[6,27–30] than for nanotube walls[27,31–33]. Moreover, graphene layers can rotate relative to each other to incommensurate states (see FIG. 1a and b) in which the potential relief of the interlayer interaction energy is smooth. At relative rotation angles $\varphi_0 = 0°, 60°, 120°$ etc., graphene layers are commensurate (see FIG. 1a) and the potential energy relief has significant barriers to the relative motion of the layers. However, when the layers are rotated relative to each other by an angle $\varphi_0 + \delta\varphi < \varphi < \varphi_0 - \delta\varphi + 60°$ ($\delta\varphi \approx a_0 / L$, where $a_0$ is the lattice constant of graphene and $L$ is the size of the layers), the layers become incommensurate (see FIG. 1b) and the barriers to the relative motion of the layers disappear[28,34–36]. Such incommensurate states are observed in the form of so-called Moiré patterns[37–39]. Though the transition of graphene layers to the incommensurate states has a considerable energy cost[29,30], it was shown to play a significant role in the static friction[34–36] and thermally-activated diffusion[29,30] of graphene flakes on graphene layers. Here we show that this phenomenon also has implications for the telescopic motion of graphene flakes.

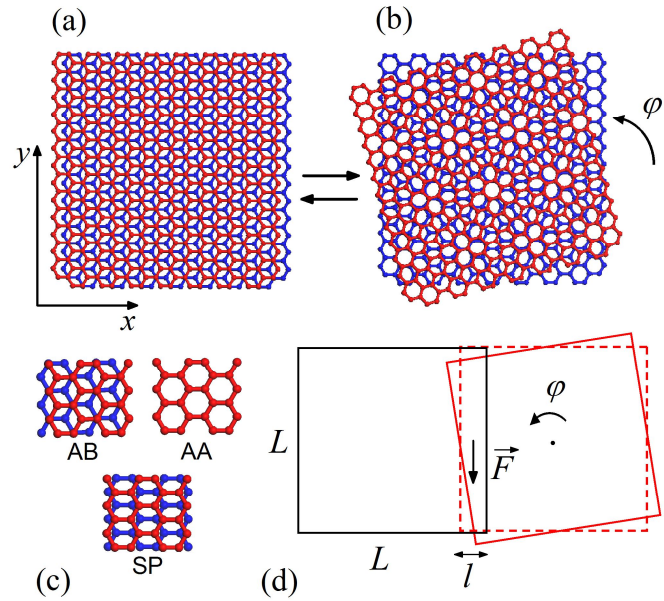

FIG. 1. (a) Commensurate ($\varphi = 0°$) and (b) incommensurate ($\varphi = 15°$) states of graphene flakes. (c) Stackings of commensurate graphene layers. (d) Schematic representation of the rotation of the extended graphene flake around its center of mass induced by the corrugation of the potential energy relief.

Therefore, investigation of the self-retracting motion of graphene flakes is of interest both for elaboration of graphene-based NEMS and for study of basic tribological properties of matter at the nanometer scale. In the present paper we examine the self-retracting motion of graphene flakes by molecular dynamics (MD) simulations. The analysis of the simulation results shows that the differences in the potential energy reliefs for graphene layers and nanotubes walls listed above lead to essentially different dynamic behavior of these systems. First, telescopic oscillations are suppressed for graphene flakes due to the excitation of flexural vibrations of the flakes. Second, the self-retracting translational motion of the extended flake is often accompanied by the rotation of the flake. The influence of temperature, the direction of the telescopic extension and the initial orientation and position of the extended flake on the possibility and characteristics of the self-retracting motion of the flake is investigated.

The paper is organized in the following way. In Sec. II we analyze the potential energy relief corresponding to the telescopic motion of one graphene flake on another. In Sec. III we describe the model considered for the MD simulations and discuss the results of the MD simulations. Our conclusions are summarized in Sec. IV.

## II. ANALYSIS OF POTENTIAL ENERGY RELIEF

Since the relative motion of graphene flakes is determined by the potential relief of their interaction energy, let us first analyze the potential energy relief corresponding to the telescopic extension of one graphene flake on another. We have calculated the potential energy relief for graphene flakes of 34 Å x 34 Å size at the equilibrium distance between the flakes of 3.4 Å. The van der Waals interaction between atoms of the flakes is described using the potential developed recently on the basis of calculations in the framework of the dispersion-corrected density functional theory[6,7]. This potential was shown to adequately describe the properties of graphite and the small relative vibrations of graphene layers[6,7]. The cutoff distance of the van der Waals potential is taken equal to 25 Å. The covalent carbon-carbon interactions in the flakes are described by the Brenner potential[40]. To calculate the dependences of the interaction energy of the graphene flakes on their relative position and orientation, the structures of the flakes are separately relaxed using the Brenner potential and then the upper flake is rigidly shifted and rotated parallel to the underlying flake.

Before proceeding with the discussion of the calculated potential energy relief for the graphene flakes let us remind the characteristics of the potential energy relief for a graphene flake on a periodic graphene layer[6,7,29,30,34–36]. The interaction energy between the graphene flake and the graphene layer can be considered as a sum of two components: a constant negative binding energy $-\varepsilon_0$ = –46.9 meV/atom (for the potential[6,7] used in the present paper) and a positive corrugation describing the dependence of the interaction energy on the stacking of the flake and the layer and equal to zero at minima of the potential energy relief. Maxima and minima of the potential energy relief of the flake commensurate with the underlying graphene layer correspond to the AA and AB stackings (see FIG. 1c), respectively. The energy difference between these states is found to be $\varepsilon_{max}$ = 19.4 meV/atom. The barriers for transitions of the graphene flake between adjacent energy minima correspond to the saddle point (SP) stacking (see FIG. 1c). The magnitude of these barriers is $\varepsilon_{com}$ = 2.1 meV/atom. With the relative rotation of the flake, the magnitude of corrugation of the potential energy relief decreases and becomes negligibly small at rotation angles exceeding $\delta\varphi \approx a_0/L$, where $a_0 = 2.46$ Å is the lattice constant of graphene and $L$ is the size of the flake. The energy required for the relative rotation of the graphene flake to the incommensurate states is $\varepsilon_{in}$ = 6.4 meV/atom. The total binding energy of the graphene flake on the periodic graphene layer and the magnitude of corrugation of the potential energy relief are proportional to the overlap area between the flake and the layer and are constant at any displacement of the flake relative to the layer.

Let us now discuss the potential energy relief calculated for the telescopically extended graphene flakes (see FIG. 2). Analogous to the graphene flake on the periodic graphene layer[29,30,34–36], the interaction energy between the graphene flakes can roughly be considered as a sum of the binding energy and the corrugation describing the dependence of the interaction energy on the stacking of the flakes. However, different from the case of the graphene flake on the periodic graphene layer, the overlap area between the flakes changes with the relative displacement of the extended flake, providing the dependence of the binding energy and the magnitude of corrugation on the relative position of the flakes (see FIG. 2a). The dependence of the binding energy on the overlap area of the flakes is the cause of the self-retracting motion of the extended flake. To analyze the dependence of the barrier to the relative motion of the flakes on their relative position we introduce the barrier $E_{\text{bar}}$ to the self-retracting motion of the commensurate flake. The barrier $E_{\text{bar}}$ is defined as the energy difference between the local energy minimum around which the flake is placed initially and the nearest saddle point in the direction opposite to the direction of extension. This barrier can be estimated as

$$E_{\text{bar}} = \frac{1}{\sigma}\left( \varepsilon_{\text{com}} lL - \varepsilon_0 \frac{a_0}{2\sqrt{3}} L \cos\alpha \right), \tag{1}$$

where $l$ is the overlap length of the flakes in the direction of extension, $\alpha$ is the angle between the direction from the energy minimum to the saddle point and the direction opposite to the direction of extension and $\sigma$ is the area per atom in graphene. The first term in this formula is related to an increase of the magnitude of corrugation of the potential energy relief with an increase of the overlap area of the flakes and is independent of the direction of extension. The second term in Eq. (1) is related to an increase of the binding energy of the flakes in magnitude with the retraction of the extended flake. This term does not depend on the overlap length of the flakes but depends on the direction of extension. For the case of extension along the zigzag direction, the angle $\alpha = 30°$ for any energy minima around which the flake is placed initially. For the flake extended along the armchair direction, the energy minima are not equivalent. The angle $\alpha$ takes the values $\alpha = 60°$ and $0°$ for displacements from the energy minima denoted by AB1 and AB2 in FIG. 3a to the saddle points denoted by SP1 and SP2, respectively. As follows from Eq. (1), the dependences of the barrier $E_{\text{bar}}$ to the self-retracting motion of the flake on the overlap length $l$ are shifted relative to each other for different directions of extension. With increasing the overlap length of the flakes $l \gg a_0$, the barrier to the self-retracting motion of the flake tends to $E_{bar} = \varepsilon_{com} lL / \sigma$ = 0.8 meV/Å$^2$·$lL$, which is the barrier to the relative motion of the graphene flake of size $l$ x $L$ on the periodic graphene layer.

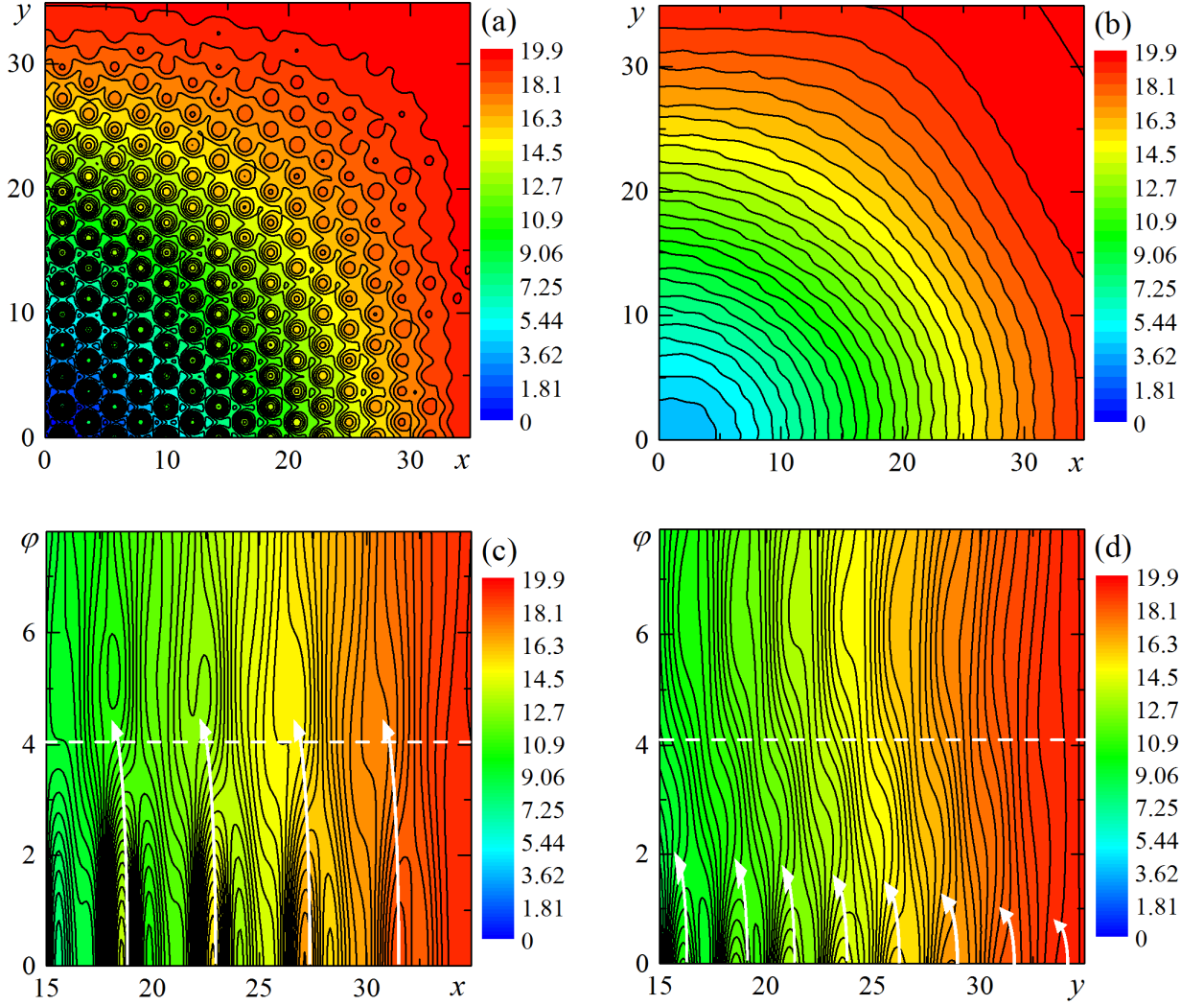

FIG. 2. Calculated interaction energy (in eV) of the graphene flakes of 34 Å x 34 Å size (446 atoms) at the equilibrium interlayer distance 3.4 Å as a function of the relative position $x$, $y$ (in Å; axes $x$ and $y$ are chosen along the armchair and zigzag directions, respectively) of the center of mass of the extended flake and the rotation angle $\varphi$ (in degrees) of the flake around its center of mass: (a) $\varphi = 0°$, (b) $\varphi = 30°$, (c) $y = 0$ and (d) $x = 0$. The equipotential lines are drawn with the step of (a, b) 0.6 eV and (c, d) 0.2 eV. The energy is given relative to the global energy minimum. The dashed lines in figures (c, d) correspond to the critical rotation angle $\delta\varphi = a_0 / L$ above which the extended flake is assumed to be incommensurate. The arrows in figures (c, d) demonstrate the energy favorable rotation of the extended flake.

By calculations of the energy difference between the SP and AB stackings, we have confirmed that the barrier $E_{bar}/L$ to the self-retracting motion of the flake per unit width of the flake in the direction perpendicular to the direction of extension depends linearly on the overlap length $l$ of the flakes with the same coefficient 0.8 meV/Å$^2$ for the armchair and zigzag directions (see FIG. 4). However, as the dependences are shifted, the barriers are different for the cases of extension along the armchair and zigzag directions. There is no barrier to the self-retracting motion of the flake extended along the zigzag direction at overlap lengths less than 10 Å. For the flake extended along the armchair direction there is also no barrier to the self-retracting motion of the flake at overlap lengths below 15 Å for the energy minima denoted by AB2 in FIG. 3a, whereas for the energy minima denoted by AB1, this barrier is non-zero for any overlap length. For example, the barrier for an AB1 energy minimum is about 0.08 eV at the extension of the considered flake by 30 Å along the armchair direction. Such a barrier can hinder the start of the self-retracting motion of the flake even at room temperature.

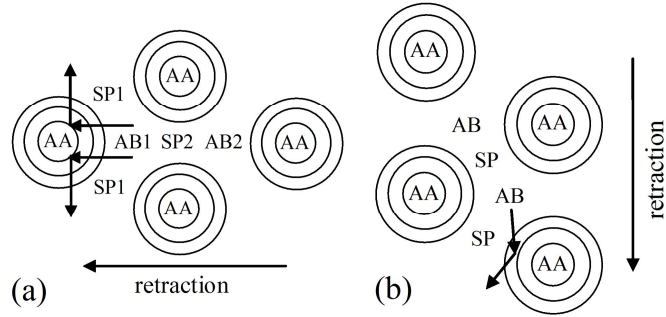

FIG. 3. Schematic representation of scattering on hills of the potential energy relief for the graphene flake extended along the (a) armchair and (b) zigzag directions. The energy minima are denoted as "AB", "AB1" and "AB2". The energy hills are denoted as "AA". The saddle points are denoted as "SP", "SP1" and "SP2". The direction of motion of the flake is indicated by the arrows.

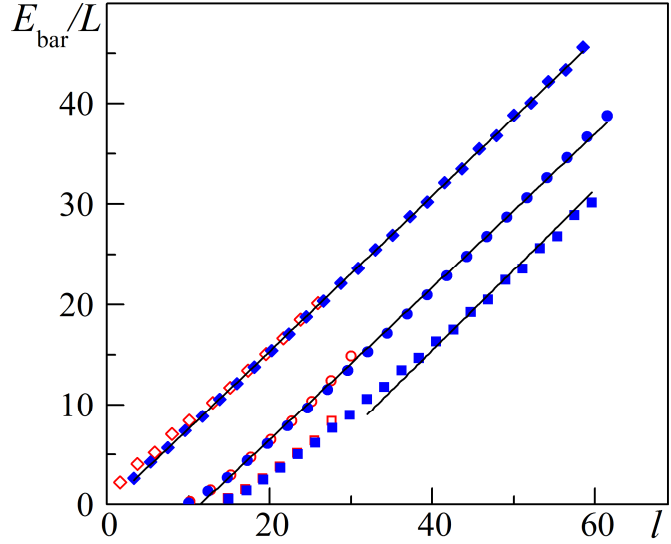

FIG. 4. Calculated barrier $E_{\text{bar}}/L$ (in eV/Å) to the self-retracting motion of the extended flake per unit width of the flake in the direction perpendicular to the direction of extension as a function of the overlap length $l$ (in Å) of the flake in the direction of extension. The data for the armchair direction are shown as diamonds and squares for the AB1 and AB2 minima (see FIG. 3a), respectively. The results for the zigzag direction are shown with the circles. The open and filled symbols correspond to the flakes of 34 Å x 34 Å and 68 Å x 68 Å size, respectively. The solid lines with the slope 0.8 meV/Å² show linear approximations of the obtained data.

In the present paper we perform MD simulations of the self-retracting motion of the extended flake for initial overlap lengths below 10 Å. For these overlap lengths, the start of the self-retracting motion of the flake is energetically favorable for any initial position of the flake extended along the zigzag direction, while for the flake extended along the armchair direction there are states around the AB1 energy minima for which the start of the self-retracting motion is possible only in the result of thermodynamic fluctuations. Therefore, the dynamic behavior of the extended flake at these initial overlap lengths is expected to be different for the armchair and zigzag directions of extension.

Upon the relative rotation of the graphene flakes (see FIG. 1b), the potential energy relief of the flakes becomes smooth (see FIG. 2b). This should facilitate the retraction of the extended flake initially rotated to an incommensurate state, which is confirmed for flakes of μm size by the experiment[3] and for flakes of nm size by our MD simulations (see Sec. III). Moreover, our simulations (see Sec. III) reveal that the rotation of the extended flake from the commensurate to incommensurate states often takes place during the self-retracting motion of the flake. To analyze the reasons of this rotation let us consider the interaction energy of the flakes as a function of the rotation angle of the extended flake (see FIG. 2c and d). It is seen from FIG. 2c and d that on the hills of the potential energy relief the extended flake experiences a torque which induces the rotation of the flake (as indicated by the arrows). The extended flake can be initially placed on a potential energy hill or the flake can start the self-retracting motion and acquire the kinetic energy sufficient for climbing the potential energy hills on the way of retraction of the flake.

As opposed to the case of a graphene flake on a periodic graphene layer[29,30], the telescopically extended flake experiences a non-zero torque even at the zero rotation angle. The origin of this additional torque which leads to the start of the rotation of the extended flake with the initial commensurate orientation can be explained as follows. The force of the interaction between the flakes is applied only to the atoms of the extended flake located in the overlap area of the flakes (see FIG. 1d). Therefore, the force acting on the flake on slopes of the potential energy hills in the direction perpendicular to the direction of extension exerts the torque inducing the rotation of the flake around its center of mass.

Note that the described mechanism of rotation of the extended flake to the incommensurate states can take place only for the flake which is able to start the self-retracting motion. The rotation of the commensurate flake locked near a local energy minimum is possible only due to thermodynamic fluctuations, similar to the thermally-activated diffusion of the graphene flake on the periodic graphene layer[29,30]. The thermally-activated rotation takes a long time and is of minor importance for the operation of graphene-based NEMS in which the fast retraction of the extended flake is desirable.

The patterns of equipotential lines in FIG. 2c and d demonstrate that the rotation of the flake to the incommensurate states by the angle $\delta\varphi$ is energetically favorable for the flake extended along the armchair direction, while for the flake extended along the zigzag direction, the rotation by angles smaller than $\delta\varphi$ is expected. Thus, the probability for the extended flake to rotate to the incommensurate states should depend on the direction of extension. This can be explained as follows. For the flake extended along the armchair direction, the potential energy hills lie directly on the way of retraction of the flake (see FIG. 3a). In this case, the chances for the flake to get on slopes of the potential energy hills where the flake experiences the torque sufficient for the rotation to the incommensurate states are high. For the flake extended along the zigzag direction, the potential energy hills lie aside the way of retraction of the flake (see FIG. 3a) and the probability for the flake to experience the torque sufficient for the rotation to the incommensurate state should be lower than in the case of extension along the armchair direction.

On the basis of the above analysis of the potential energy relief, the following features of the self-retracting motion of the flake can be predicted. First, due to the significant variation of the interaction energy of the commensurate flakes at the scale of the lattice constant of graphene, the behavior of the extended flake with the initial commensurate orientation should be highly sensitive to the initial position of the flake on the subatomic scale. Second, the rotation of the flake to the incommensurate states should facilitate the retraction of the flake. Third, different behavior of the flake is expected for the cases of extension of the flake along the armchair and zigzag directions due to the differences in the barriers to the self-retraction motion and the probabilities of rotation to the incommensurate states. To check these predictions we have performed the MD simulations of the self-retracting motion of the graphene flake.

## III. MOLECULAR DYNAMICS SIMULATIONS

The MD simulations of the self-retracting motion of a graphene flake are performed for the system consisting of three graphene flakes of 34 Å x 34 Å size at the equilibrium distance of 3.4

Å from each other (see FIG. 1 and FIG. 5). The in-house MD-kMC[41] (Molecular Dynamics – kinetic Monte Carlo) code is used. The integration time step is 0.6 fs. The temperature of the middle flake is maintained by rescaling atomic velocities every 0.1 ps (the Berendsen thermostat[42]). All atoms of the bottom flake are fixed. At the beginning of simulations, the upper flake is extended telescopically by about 30 Å. The extended flakes at different initial positions within the unit cell of graphene, with initial orientations corresponding to the commensurate ($\varphi = 0°$) and incommensurate states ($\varphi = 30°$), displaced along the armchair and zigzag directions, at temperatures 4 and 300 K are considered.

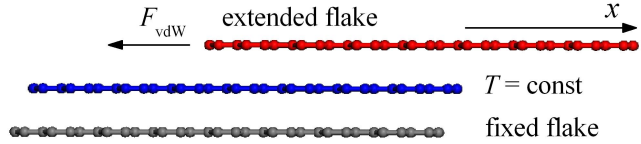

FIG. 5. Schematic representation of the telescopic motion of a graphene flake (the system under consideration).

The simulations show that the self-retracting motion of the extended flake with the initial commensurate orientation proceeds in diverse ways even for nearly the same initial conditions (different only in the initial distribution of thermal velocities of the atoms and/or in the initial position of the extended flake within the unit cell of graphene, see FIG. 6 and supplemental movies[43]). In most cases, retraction of the extended flake occurs. The self-retracting motion of the flake can be fast (at times ~ 10 ps) or slow (at times ~ 50 ps). Sometimes the flake performs 1 – 2 damped oscillations. Rarely there is also a time lag for retraction of the flake. In some cases, no retraction at all is observed at simulation times of hundreds of picoseconds. This diversity of the dynamic behavior of the graphene flake is provided by thermodynamic fluctuations, similar to NEMS based on carbon nanotubes[23,25] as well as by the properties of the potential energy relief of the flake discussed in the previous section.

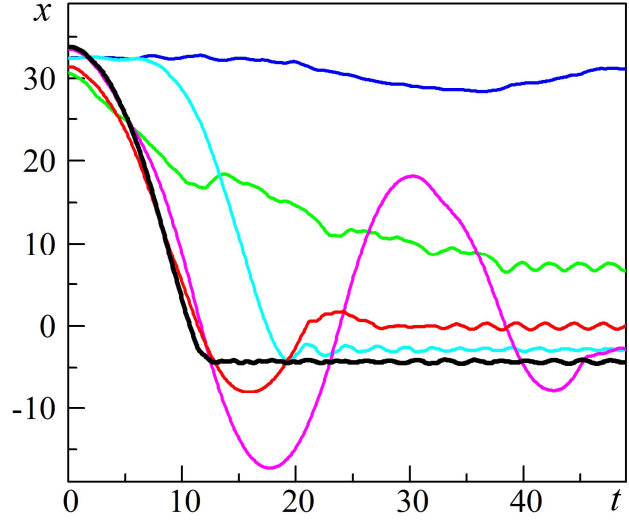

FIG. 6. Different types of calculated dependences of displacement $x$ (in Å) of the flake with the initial commensurate orientation extended in the armchair direction on time $t$ (in ps) at temperature 300 K. The coordinate $x = 0$ corresponds to the ground state position of the flake before extension.

As expected from the potential energy relief (see Sec. II), the behavior of the flake which is initially in the commensurate state is found to be strongly sensitive to the initial position of the flake at the scale of the lattice constant of graphene. For example, the initial positions of the commensurate flake extended by 30, 31 and 32 Å in the armchair direction correspond to the positions near an AB2 energy minimum (see FIG. 3a) and slopes of a potential energy hill with attractive and repulsive forces, respectively. For each of these initial positions, we have performed 10 simulations of 50 ps duration at temperature 4 K. The complete retraction of the flake is observed in 0, 9 and 2 simulations, respectively. However, experimentally it can be difficult to control the initial position of the flake with such a high accuracy. Therefore, the results of calculations for a particular initial extension of the flake in the commensurate state have been averaged over the unit cell of graphene.

To characterize the diversity of the dynamic behavior of the flake we determine the fractions $\chi$ and $\zeta$ of simulations in which the rotation of the flake to the incommensurate states and $1 - 2$ damped oscillations of the flake are observed, respectively. In addition to these phenomena, the diversity of the flake behavior is related to the different duration of motion after crossing the position corresponding to the zero displacement. In several simulations at 300 K with the initial extension of the flake of 30 Å along the armchair direction, a time lag of 5 – 30 ps for the start of the self-retracting motion of the flake is also seen. To give an account of these observations two average times are calculated: the average time $\tau_r$ of retraction of the flake (the time in which the relative displacement of the flake along the direction of extension changes from the initial value to zero or to the final value if the flake stops before it crosses the position corresponding to the zero displacement) and the average time $\tau_s$ of complete stop of the flake. The average coordinates $x$, $y$ of the final position of the extended graphene flake and root-mean-square deviations $\sigma_x$, $\sigma_y$ of the final coordinates of the flake are also found. These data obtained on the basis of the MD simulations are given in Table I.

The simulations for the extended graphene flake with the initial commensurate orientation show almost no telescopic oscillations of the extended flake ($\zeta \ll 1$ in Table I). Furthermore, it is found that the self-retracting motion of the flake is often accompanied by the rotation to the incommensurate states ($\chi \sim 1$ in Table I, see also supplemental movies[43] and FIG. 7). As seen from FIG. 7, the rotation of the flake proceeds simultaneously with the displacement of the flake in the direction perpendicular to the direction of extension. This supports the hypothesis that the rotation is induced by the torque of the force acting on the flake on steep slopes of the potential energy hills in the direction perpendicular to the direction of extension. The rotation of the flake to the incommensurate states on the potential energy hills is also indicated in FIG. 8 for trajectories of the flake. It follows from FIG. 8 that the rotation of the flake occurs as a result of getting off the potential energy hill on which the flake is placed initially or due to scattering on the potential energy hills on the way of retraction of the flake.

Table I. Average final position $x$, $y$ (in Å) of the extended graphene flake relative to the global energy minimum, root-mean-square deviations $\sigma_x$, $\sigma_y$ (in Å) of the final coordinates of the flake, average time $\tau_r$ (in ps) of retraction of the flake, average time $\tau_s$ (in ps) of complete stop of the flake and fractions $\chi$ and $\zeta$ of simulations in which the rotation of the flake to the incommensurate states and several oscillations of the flake are observed, respectively, for different temperatures $T$ (in K), initial rotation angles $\varphi$ (in degrees) of the extended flake and directions of the initial extension of the flake. The results are obtained on the basis of 25 – 35 simulations of 50 ps duration with the initial position of the flake within a unit cell of graphene at the distance of about 30 Å from the global energy minimum, respectively. The axes $x$ and $y$ are chosen along the armchair and zigzag directions, respectively.

| $T$ | Extension | $\varphi$ | $\tau_r$ | $\tau_s$ | $\zeta$ | $\chi$ | Final position | | | | | | | |
|---|---|---|---|---|---|---|---|---|---|---|---|---|---|---|
| | | | | | | | with rotation | | | | without rotation | | | |
| | | | | | | | $x$ | $y$ | $\sigma_x$ | $\sigma_y$ | $x$ | $y$ | $\sigma_x$ | $\sigma_y$ |
| 4 | $x$ | 0 | 14.1 | 21.3 | 0.06 | 0.80 | 2.7 | 0.3 | 6.3 | 1.8 | 28.7 | 0.8 | 4.6 | 0.6 |
| 300 | $x$ | 0 | 15.2 | 23.5 | 0.04 | 0.85 | 0.8 | -0.3 | 4.7 | 1.8 | 24.7 | 0.5 | 6.0 | 0.4 |
| 4 | $y$ | 0 | 11.5 | 13.4 | <0.03 | 0.37 | 2.0 | -1.3 | 1.6 | 3.3 | 1.6 | 2.1 | 1.1 | 6.7 |
| 300 | $y$ | 0 | 12.2 | 26.3 | 0.12 | 0.88 | -0.2 | -1.5 | 1.7 | 4.0 | 0.1 | -2.8 | 1.2 | 1.9 |
| 4 | $x$ | 30 | 9.5 | 25.0 | 0.14 | 1 | -5.5 | 0.0 | 8.9 | 1.2 | | | | |
| 300 | $x$ | 30 | 9.6 | 33.5 | 0.26 | 1 | -0.6 | 0.1 | 7.6 | 2.4 | | | | |

The commensurate flake extended along the armchair direction is found either to retract poorly or to rotate to the incommensurate states and to retract completely (see Table I). In 80 and 85% of the simulations at temperatures 4 and 300 K, respectively, the flake starts the self-retracting motion and rotates to the incommensurate states. In 17 and 12% of the simulations the flake is observed to be trapped in the initial local energy minimum. In 3 and 4 % the flake starts the self-retracting motion but gets trapped in another energy minimum due to the energy dissipation. The fractions listed above weakly depend on temperature, indicating that the probabilities for the flake to start the self-retracting motion and to rotate to the incommensurate states are determined by the initial position and energy of the flake rather than by thermodynamic fluctuations. FIG. 8a shows that most of the trajectories of the flake extended along the armchair direction go directly through the potential energy hills on which the flake experiences scattering accompanied by the rotation to the incommensurate states.

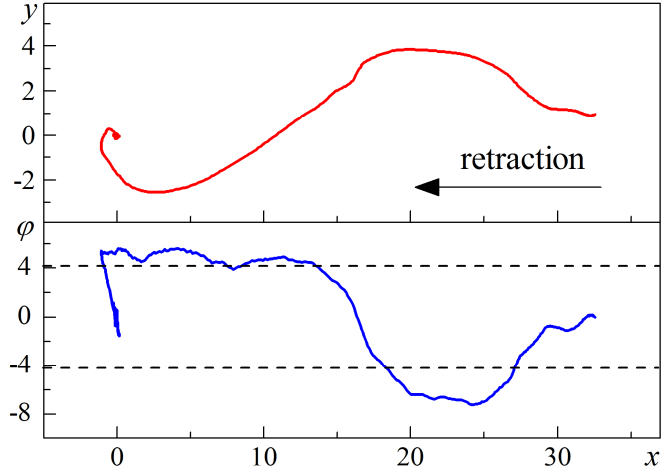

FIG. 7. Calculated displacement $y$ (in Å) of the extended graphene flake in the zigzag direction (the direction perpendicular to the direction of extension) and rotation angle $\varphi$ (in degrees) of the flake around the center of mass as functions of the displacement $x$ (in Å) of the flake in the armchair direction (the direction of extension) at temperature 4 K. The coordinate $x = 0$ corresponds to the ground state position of the flake. The angle $\varphi = 0$ corresponds to the commensurate state. The critical angles $\pm\delta\varphi$ at which the flake becomes incommensurate are shown with the dashed lines.

As shown in the previous section, for the flake extended along the zigzag direction there is no barrier to the self-retracting motion for overlap lengths below 10 Å (see FIG. 4). In this case, the complete retraction of the flake is observed in all simulations independent of the rotation to the incommensurate states (see Table I). The fraction $\chi$ of simulations in which the flake rotates to the incommensurate states is relatively small at 4 K and increases with temperature (see Table I). The analysis performed in the previous section showed that the potential energy relief of the flake extended along the zigzag direction favors the rotation of the flake only to the angles below $\delta\varphi \approx a_0 / L$, which are insufficient to completely eliminate the corrugation of the potential energy relief (see FIG. 2d). In agreement with this conclusion, most of the trajectories of the flake extended along the zigzag direction at temperature 4 K lie around the potential energy hills and the rotation angle of the flake does not reach the critical angle $\delta\varphi$ at which the flake becomes incommensurate with the other flakes (see FIG. 8b). However, with increasing the temperature, the rotation of the flake to angles above $\delta\varphi$ becomes possible.

The extended flake which is initially rotated by 30º to the incommensurate state is found to retract in all simulations (see the data in Table I for $\varphi = 30°$). The fraction of simulations is which several telescopic oscillations of the flake are observed increases compared to the case of the initial commensurate orientation of the flake (see Table I). However, this fraction is still much less than unity. This means that the telescopic motion of graphene flakes cannot be used in oscillators proposed in Ref. 3.

The initial orientation of the flake also determines the time of retraction. If the extended flake is initially in the commensurate state, the corrugation of the potential energy relief slows down the self-retracting motion of the flake. On the contrary, for the flake which is initially rotated to the incommensurate state, there are no barriers to the motion of the flake and its retraction occurs as fast as possible with the smallest average time $\tau_r$ of retraction (see Table I).

The final position of the retracted flake does not always correspond to the global energy minimum. The flake can stop in numerous metastable states on the potential energy relief. The dispersion of the final coordinate of the flake in the direction of extension has a significant value of 4 – 10 Å for all the considered cases (see Table I). This dispersion is somewhat higher for the flake extended along the armchair direction than for the flake extended along the zigzag direction. The dispersion further increases when the flake is initially rotated to the incommensurate state.

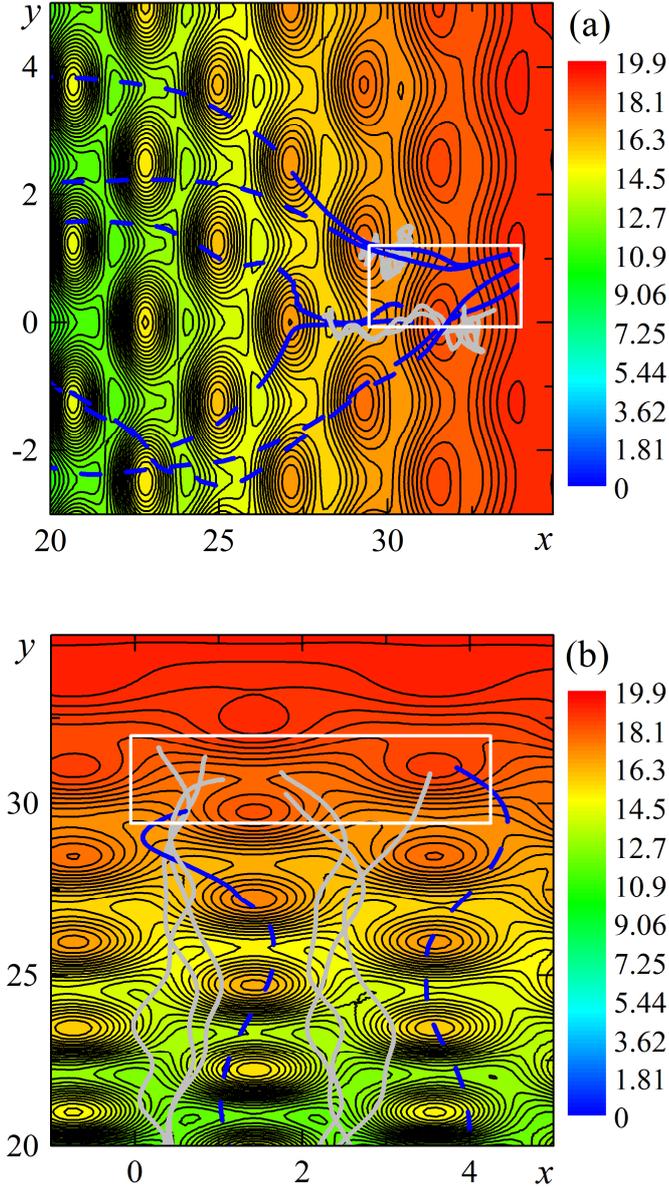

FIG. 8. Calculated trajectories of the flake extended along the (a) armchair and (b) zigzag directions at temperature 4 K. The trajectories for the flake staying in the commensurate state are shown in gray. The trajectories in which the rotation of the flake to the incommensurate states is observed are shown in blue. The solid and dashed lines correspond to the commensurate and incommensurate states of the flake. The calculated interaction energy (in eV) of the graphene flakes of 34 Å x 34 Å size in the commensurate state ($\varphi = 0°$) is given at the equilibrium interlayer distance as a function of the relative position $x$, $y$ (in Å; axes $x$ and $y$ are chosen along the armchair and zigzag directions, respectively) of the center of mass of the extended flake. The equipotential lines are drawn with a step of 0.2 eV. The white rectangular shows the unit cell of graphene in which the initial positions of the flake are distributed.

To clarify the role of rotation of the extended flake to the incommensurate states we have performed the MD simulations for the flakes with the fixed commensurate and incommensurate orientations extended along the armchair direction. For the flake with the fixed commensurate orientation, no telescopic oscillations are observed. The average final position of such a flake in the armchair direction is found to be 15.5 Å at 300 K and 21.2 Å at 4 K based on ten simulations at each temperature. The flake with the fixed incommensurate orientation (rotated by 30$^\text{o}$ with respect to the commensurate orientation) performs telescopic oscillations with the Q-factor $Q = 2.00 \pm 0.08$ at 300 K and $Q = 2.25 \pm 0.07$ at 4 K. The time of retraction of this flake corresponding to one quarter of the oscillation is 9.5 ps. These results confirm that the rotation of the flake to the incommensurate states facilitates the retraction of the flake.

Since the visual analysis of the simulation results shows a high excitation of flexural vibrations of the graphene flakes, such a way of dissipation seems to be dominant for the telescopic motion of the extended flake. To check this assumption we have performed the simulations of the telescopic oscillations of the graphene flake with the fixed incommensurate orientation which is also constrained in the plane. In this case, the Q-factor of the oscillations along the armchair direction at temperature 300 K is found to increase up to $Q \sim 10$. Therefore, out-of-plane vibrations of atoms of the graphene flake considerably increase the dissipation of the kinetic energy of the flake. The Q-factors calculated for the graphene flake are orders of magnitude smaller than the Q-factors $Q \sim 100 - 1000$ for the telescopic oscillations of carbon nanotube walls[23,25]. This can be explained by the fact that carbon nanotubes are stiffer, and no significant flexural vibrations are detected in typical simulations of the telescopic oscillations of nanotube walls[10–26].

The results obtained in the present paper can be extended to the case of graphene flakes of μm size studied experimentally[3]. For flakes of size $L \sim 1$ μm, the critical rotation angle $\delta\varphi$ at which the flakes become incommensurate with the other flakes is very small $\delta\varphi \sim a_0 / L \sim 10^{-4}$ (see Ref. 29, 30, 35, 36). Therefore it could be difficult to detect such a small angle in experiments. We believe that most of the extended flakes in the experiment[3] were initially rotated to incommensurate states. This explains why the retraction was observed[3] with the 100% probability for the flakes of 1–2 μm size. For larger flakes, the probability of the retraction decreased, which could be attributed to non-elastic deformations of these flakes during the pull-out[3]. Assuming that the time of retraction of the flake scales with the size of the flake as $\tau_r \sim \sqrt{ms}$ (see Ref. 25), where $s$ is the length of the initial extension and $m$ is the mass of the extended flake, and using the data from Table I, we estimate the time of retraction for flakes of μm size to be of the order of microseconds, in agreement with the observations of paper[3].

## IV. CONCLUSION

We have performed the MD simulations of the self-retracting motion of a graphene flake telescopically extended from its equilibrium position on a stack of graphene flakes. It is shown that almost no telescopic oscillations of the graphene flake occur due to locking of the flake in the commensurate states with the high corrugation of the potential energy relief and the high dynamic friction force related to the excitation of flexural vibrations of the flake both in the commensurate and incommensurate states. It is demonstrated that the potential energy relief becomes smooth upon the rotation of the extended flake to the incommensurate states, which facilitates the self-retracting motion of such a flake. The rotation of the extended flake from the commensurate to incommensurate states is often observed in the MD simulations and is shown to be induced by the torque acting on the flake on potential energy hills. The flake is found to stop in numerous metastable states on the potential energy relief with the dispersion of the final position up to 10 Å for the considered system.

Mechanical oscillations (followed by the oscillation of the tunneling current) arising after switching on were shown to be the main problem which restricts the response speed of memory cells based on the bending[44] and telescopic motion[45] of carbon nanotubes. The observed ultrafast damping of the telescopic oscillations of the graphene flake makes us expect the absence of mechanical oscillations after switching on for memory cells based on the telescoping motion of an extended graphene flake. Thus, graphene is a promising material for the use in fast-responding memory cells.

**ACKNOWLEDGEMENTS**

This work has been partially supported by the RFBR grants 11-02-00604 and 10-02-90021-Bel. The calculations are performed on the SKIF MSU Chebyshev supercomputer, the MVS-100K supercomputer at the Joint Supercomputer Center of the Russian Academy of Sciences and the supercomputer at National Research Centre "Kurchatov Institute".

29. I. V. Lebedeva, A. A. Knizhnik, A. M. Popov, O. V. Ershova, Yu. E. Lozovik, and B. V. Potapkin, Phys. Rev. B 82, 155460 (2010).

30. I. V. Lebedeva, A. A. Knizhnik, A. M. Popov, O. V. Ershova, Yu. E. Lozovik, and B.V. Potapkin, J. Chem. Phys. 134, 104505 (2011).

31. E. Bichoutskaia, A. M. Popov, A. El-Barbary, M. I. Heggie, and Yu. E. Lozovik, Phys. Rev. B 71, 113403 (2005).

32. E. Bichoutskaia, M. I. Heggie, A. M. Popov, and Yu. E. Lozovik, Phys. Rev. B 73, 045435 (2006).

33. J.-C. Charlier and J.-P. Michenaud, Phys. Rev. Lett. 70, 1858 (1993).

34. N. Sasaki, K. Kobayashi and M. Tsukada, Phys. Rev. B 54, 2138 (1996).

35. M. Dienwiebel, G. S. Verhoeven, N. Pradeep, J. W. M. Frenken, J. A. Heimberg, and H. W. Zandbergen, Phys. Rev. Lett. 92, 126101 (2004).

36. A. E. Filippov, M. Dienwiebel, J. W. M. Frenken, J. Klafter, and M. Urbakh, Phys. Rev. Lett. 100, 046102 (2008).

37. J. H. Warner, M. H. Rummeli, T. Gemming, B. Buchner, and G. A. D. Briggs, Nano Lett. 9, 102 (2009).

38. Z. Y. Rong and P. Kuiper, Phys. Rev. B 48, 17427 (1993).

39. Y. Gan, W. Chu, L. Qiao, Surf. Sci. 539, 120 (2003).

40. D. W. Brenner, O. A. Shenderova, J. A. Harrison, S. J. Stuart, B. Ni and S. B. Sinnott, J. Phys.: Condens. Matter 14, 783 (2002).

41. http://www.kintechlab.com/products/md-kmc/

42. H. J. C. Berendsen, J. P. M. Postma, W. F. van Gunsteren, A. DiNola, and J. R. Haak, J. Chem. Phys. 81, 3684 (1984).

43. See Supplemental Material at http://link.aps.org/supplemental/10.1103/PhysRevB.84.245437 for movies.
23

44. J. W. Kang, O. K. Kwon, J. H. Lee, H. J. Lee, Y.-J. Song, Y.-S. Yoon, and H. J. Hwang, Physica E 33, 41 (2006).

45. J. W. Kang and Q. Jiang, Nanotechnology 18, 095705 (2007).24